\documentclass[aps,preprint]{revtex4}
\usepackage{amssymb,epsf}
\usepackage{amsmath}
\usepackage{graphicx}
\begin{document}
\title{Static and Radiating Solutions of Lovelock Gravity in the Presence of a Perfect Fluid}
\author{M. H. Dehghani$^{1,2}$\footnote{email address:
mhd@shirazu.ac.ir} and N. Farhangkhah$^{1}$}
\affiliation{$^1$ Physics Department and Biruni Observatory, College of Sciences, Shiraz University, Shiraz 71454, Iran\\
        $^2$ Research Institute for Astrophysics and Astronomy of Maragha (RIAAM), Maragha,
        Iran}

\begin{abstract}
We present a general solution of third order Lovelock gravity in the
presence of a specific type II perfect fluid. This solution
for linear equation of state, $p=w(\rho-4B)$ contains all the known
solutions of third order Lovelock gravity in the literature and
some new static and radiating solutions for different values of
$w$ and $B$. Specially, we consider the properties of static
and radiating solutions for $w=0$ and $w=(n-2)^{-1}$ with
$B=0$ and $B\neq0$. These solutions are asymptotically flat for
$B=0$, while they are asymptotically (anti)-de Sitter for $B\neq0$.
The new static solutions for these choices of $B$ and $w$ present black holes with
one or two horizons, extreme black holes or naked singularities provided the parameters
of the solutions are chosen suitable. The static solution with $w=0$
and vanishing geometrical mass ($m=0$) may
present a black hole with two inner and outer horizons. This is a peculiar
feature of the third order Lovelock gravity, which does not occur in
lower order Lovelock gravity. We also, investigate the properties of radiating
solutions for these values of $B$ and $w$, and compare the singularity strengths of them
with the known radiating solutions of third order Lovelock gravity.

\end{abstract}

\maketitle
\section{Introduction}

The dynamics of our universe at the classical level may be described by a
gravitational field equation in which the contribution of energy content of
the universe is represented by an energy momentum tensor appearing on the
right-hand side (RHS) of this equation. The left-hand side (LHS) represents
pure geometry given by the curvature of spacetime. Gravitational equations
in their original form with energy-momentum tensor of normal matter cannot
lead to an accelerated expanding universe, while high-precision
observational data have confirmed with startling evidence that the universe
is undergoing a phase of accelerated expansion. There are, then, two ways to
obtain accelerated expansion of the universe. The first way is by modifying
the RHS and supplementing energy-momentum tensor by dark energy component.
The second way, which one may supersede the dark energy, is the
generalization of the LHS of the field equation. A natural generalization of
LHS of the field equation with the assumption of Einstein--that it is the
most general symmetric conserved tensor containing no more than two
derivatives of the metric-- is Lovelock gravity \cite{Lov}. Also, the action
of Lovelock gravity may be viewed as the low-energy effective action of
string theory at the classical level \cite{string}. Asymptotically flat
solution of second order Lovelock gravity has been introduced in \cite{Boul}%
, while this kind of solution for third order Lovelock gravity has been
introduced in \cite{Deh1}. In recent years, solutions of Lovelock gravity in the
presence of more general matter distribution have been investigated.
Here, we want to introduce the exact solutions of third order Lovelock gravity in the
presence of a perfect fluid with linear equation of state and investigate
their properties.
This is motivated by the fact that third order Lovelock gravity has some peculiar features
which do not occur in lower order Lovelock gravity. Recently, one of us has shown that
a topological black hole of third order Lovelock gravity has an unstable
phase which does not occur in the lower order Lovelock gravity \cite{DP}. The black hole solutions
of this theory and their thermodynamics have been considered in \cite{Third},
while radiating solution of this theory in the presence of a null fluid has been
considered in \cite{Far}. Recently, the generalized Vaidya
spacetime in Lovelock gravity has been investigated in \cite{Cai1}.
Also, some efforts have been done
in the computation of the conserved quantities of third order Lovelock gravity
through the use of the counterterm method \cite{Conter}.

The outline of this work is as follows. In Sec. \ref{Gen}, we give a brief
review of field equations and introduce a general solution of third
order Lovelock gravity in the presence of a specific type II perfect fluid. Section
\ref{Stat} is devoted to the presentation of the static solutions for a
type I perfect fluid with linear equation of state. In Sec. \ref{Rad}, we
generalize these solutions to the case of radiating solutions, and study the
properties of them. Finally, we give some concluding remarks.

\section{General Solutions}\label{Gen}

The field equation of third order Lovelock \ gravity in the absence of
cosmological constant may be written as \cite{Lov}
\begin{equation}
\mathcal{G}_{\mu \nu }=\sum_{i=1}^{3}\alpha _{i}^{\prime }[\mathcal{H}_{\mu
\nu }^{(i)}-\frac{1}{2}g_{\mu \nu }\mathcal{L}^{(i)}]=\kappa_n^2 T_{\mu
\nu },  \label{field}
\end{equation}
where $\alpha _{i}^{\prime }$'s are Lovelock coefficients, $T_{\mu \nu }$ is
the energy-momentum tensor of matter field, $\mathcal{L}^{(1)}=R$, $\mathcal{%
L}^{(2)}=R_{\mu \nu \gamma \delta }R^{\mu \nu \gamma \delta }-4R_{\mu \nu
}R^{\mu \nu }+R^{2}$, and
\begin{eqnarray}
\mathcal{L}^{(3)} &=&2R^{\mu \nu \sigma \kappa }R_{\sigma \kappa \rho \tau
}R_{\phantom{\rho \tau }{\mu \nu }}^{\rho \tau }+8R_{\phantom{\mu
\nu}{\sigma \rho}}^{\mu \nu }R_{\phantom {\sigma \kappa} {\nu \tau}}^{\sigma
\kappa }R_{\phantom{\rho \tau}{ \mu \kappa}}^{\rho \tau }+24R^{\mu \nu
\sigma \kappa }R_{\sigma \kappa \nu \rho }R_{\phantom{\rho}{\mu}}^{\rho }
\notag \\
&&+3RR^{\mu \nu \sigma \kappa }R_{\sigma \kappa \mu \nu }+24R^{\mu \nu
\sigma \kappa }R_{\sigma \mu }R_{\kappa \nu }+16R^{\mu \nu }R_{\nu \sigma
}R_{\phantom{\sigma}{\mu}}^{\sigma }-12RR^{\mu \nu }R_{\mu \nu }+R^{3}
\label{L3}
\end{eqnarray}
are first, second and third order Lovelock Lagrangian, respectively. In Eq. (%
\ref{field}), $\mathcal{H}_{\mu \nu }^{(1)}$ is just the Ricci tensor, and $%
\mathcal{H}_{\mu \nu }^{(2)}$ and $\mathcal{H_{\mu \nu}}^{(3)}$ are given as
\begin{equation}
\mathcal{H}_{\mu \nu }^{(2)}=2(R_{\mu \sigma \kappa \tau }R_{\nu }^{%
\phantom{\nu}\sigma \kappa \tau }-2R_{\mu \rho \nu \sigma }R^{\rho \sigma
}-2R_{\mu \sigma }R_{\phantom{\sigma}\nu }^{\sigma }+RR_{\mu \nu }),
\label{H2}
\end{equation}

\begin{eqnarray}
\mathcal{H}_{\mu \nu }^{(3)} &=&-3(4R^{\tau \rho \sigma \kappa }R_{\sigma
\kappa \lambda \rho }R_{\phantom{\lambda }{\nu \tau \mu}}^{\lambda }-8R_{%
\phantom{\tau \rho}{\lambda \sigma}}^{\tau \rho }R_{\phantom{\sigma
\kappa}{\tau \mu}}^{\sigma \kappa }R_{\phantom{\lambda }{\nu \rho \kappa}%
}^{\lambda }+2R_{\nu }^{\phantom{\nu}{\tau \sigma \kappa}}R_{\sigma \kappa
\lambda \rho }R_{\phantom{\lambda \rho}{\tau \mu}}^{\lambda \rho }  \notag \\
&&-R^{\tau \rho \sigma \kappa }R_{\sigma \kappa \tau \rho }R_{\nu \mu }+8R_{%
\phantom{\tau}{\nu \sigma \rho}}^{\tau }R_{\phantom{\sigma \kappa}{\tau \mu}%
}^{\sigma \kappa }R_{\phantom{\rho}\kappa }^{\rho }+8R_{\phantom
{\sigma}{\nu \tau \kappa}}^{\sigma }R_{\phantom {\tau \rho}{\sigma \mu}%
}^{\tau \rho }R_{\phantom{\kappa}{\rho}}^{\kappa }  \notag \\
&&+4R_{\nu }^{\phantom{\nu}{\tau \sigma \kappa}}R_{\sigma \kappa \mu \rho
}R_{\phantom{\rho}{\tau}}^{\rho }-4R_{\nu }^{\phantom{\nu}{\tau \sigma
\kappa }}R_{\sigma \kappa \tau \rho }R_{\phantom{\rho}{\mu}}^{\rho
}+4R^{\tau \rho \sigma \kappa }R_{\sigma \kappa \tau \mu }R_{\nu \rho
}+2RR_{\nu }^{\phantom{\nu}{\kappa \tau \rho}}R_{\tau \rho \kappa \mu }
\notag \\
&&+8R_{\phantom{\tau}{\nu \mu \rho }}^{\tau }R_{\phantom{\rho}{\sigma}%
}^{\rho }R_{\phantom{\sigma}{\tau}}^{\sigma }-8R_{\phantom{\sigma}{\nu \tau
\rho }}^{\sigma }R_{\phantom{\tau}{\sigma}}^{\tau }R_{\mu }^{\rho }-8R_{%
\phantom{\tau }{\sigma \mu}}^{\tau \rho }R_{\phantom{\sigma}{\tau }}^{\sigma
}R_{\nu \rho }-4RR_{\phantom{\tau}{\nu \mu \rho }}^{\tau }R_{\phantom{\rho}%
\tau }^{\rho }  \notag \\
&&+4R^{\tau \rho }R_{\rho \tau }R_{\nu \mu }-8R_{\phantom{\tau}{\nu}}^{\tau
}R_{\tau \rho }R_{\phantom{\rho}{\mu}}^{\rho }+4RR_{\nu \rho }R_{%
\phantom{\rho}{\mu }}^{\rho }-R^{2}R_{\nu \mu }).  \label{H3}
\end{eqnarray}

Here, we consider the simplest $n$-dimensional spherically symmetric metric
with one metric function, which may be written as
\begin{equation}
ds^{2}=-f(v,r)dv^{2}+2drdv+r^{2}\gamma_{i j}d\theta^i d\theta^j,  \label{metric1}
\end{equation}
where $0\leq r<\infty $ is the radial coordinate, $-\infty <v<\infty $ is an
advanced time coordinate, and $\gamma_{i j}d\theta^i d\theta^j$ is the line element of
the $(n-2)$-dimensional unit sphere. We want to introduce the general
solutions of third order Lovelock gravity in the presence of a perfect
fluid. In Refs. \cite{Hus} and \cite{Hus-nd}, the reverse has been done in
4-dimensional and $n$-dimensional Einstein gravity, respectively. That is, a
spherically symmetric solution with a metric function $f(v,r)=1-m(v,r)/r$ is
accepted and the properties of the energy-momentum tensor have been
investigated. The energy-momentum tensor which we consider here is
the sum of two noninteracting components namely the Vaidya null radiation and
a type I perfect fluid given as \cite{Ellis}
\begin{equation}
T_{\mu \nu }\,=\sigma (v,r)v_{\mu }v_{\nu }+\rho (v,r)(v_{\mu }w_{\nu
}+w_{\mu }v_{\nu })+p(v,r)(v_{\mu }w_{\nu }+w_{\mu }v_{\nu }+g_{\mu \nu }),
\label{EMten}
\end{equation}
where $v_{\mu }=(1,0,0,...,0)$ and $w_{\nu }=(f/2,-1,0,...,0)$ are two linearly
independent future pointing null vectors in $n$ dimensions which satisfy $v_{\mu }w^{\mu }=-1$%
. The stress-energy tensor has been chosen in such a way that $T_{\mu \nu
}v^{\mu }v^{\nu }=0$ and $T_{\mu \nu }w^{\mu }w^{\nu }=\sigma $, and
therefore there is energy flux only along one of the null directions. It is
of precisely the form which gives the charged Vaidya solution of Einstein
gravity \cite{Sull} for $p=\rho$, and reduces to the energy-momentum tensor which gives
the Vaidya metric \cite{Vai} for $p=\rho =0$. The stress-energy tensor (\ref
{EMten}) satisfies the dominant or weak energy conditions if the
conditions [$\rho \geq 0$, $-p\leq \rho \leq p$ and $\sigma >0$] or [$\rho
\geq 0$, $\rho +p\geq 0$ and $\sigma >0$] are met, respectively.

Using the field equation (\ref{field}) in a unit system with $\alpha
^{\prime }=1$ and defining $\alpha _{2}^{^{\prime }}=\alpha _{2}/(n-3)(n-4)$
and $\alpha _{3}^{^{\prime }}=\alpha _{3}/3(n-3)...(n-6)$ for simplicity,
the $\mathcal{G}_{v }^{\text{ \ }v}$ and $\mathcal{G}_{v }^{\text{ \ }%
r}$\ components reduce to:
\begin{eqnarray}
&&-\frac{(n-2)}{2r^{2}}\Big\{\left[ 1+\frac{2\alpha _{2}}{r^{2}}(1-f)+\frac{%
\alpha _{3}}{r^{4}}(1-f)^{2}\right] rf^{\prime }  \notag \\
&&-(1-f)\left[ (n-3)+\frac{(n-5)\alpha _{2}}{r^{2}}(1-f)+\frac{(n-7)\alpha
_{3}}{3r^{4}}(1-f)^{2}\right] \Big\}=k_{n}^{2}\rho (v,r),  \label{rho}
\end{eqnarray}

\begin{equation}
-\frac{(n-2)}{2r}\left\{ 1+\frac{2\alpha _{2}}{r^{2}}(1-f)+\frac{\alpha _{3}%
}{r^{4}}(1-f)^{2}\right\} \dot{f}=k_{n}^{2}\sigma (v,r),  \label{sigma}
\end{equation}
where the prime and the dot denote the derivatives with respect to the
coordinates $r$ and $v$, respectively. Also, one may note that the angular
components of \ Eq. (\ref{field}) may be written as
\begin{equation}
p=-\frac{1}{(n-2)r^{n-3}}\frac{\partial}{\partial r}(r^{n-2}\rho ).  \label{p}
\end{equation}
In order to solve the field equations, we define the energy function $%
\varepsilon (v,r)$ as
\begin{equation}
\varepsilon (v,r)\equiv \frac{2k_{n}^{2}}{(n-2)}\int \rho (v,r)r^{n-2}dr.
\label{mass}
\end{equation}
Now, substituting $\rho (v,r)$ from Eq. (\ref{rho}) into Eq. (\ref{mass})
and integrating the result with respect to $r$, one obtains:
\begin{equation}
\varepsilon (v,r)=r^{n-7}\left\{ r^{4}\left[ 1-f(v,r)\right] +\alpha
_{2}r^{2}\left[ 1-f(v,r)\right] ^{2}+\frac{\alpha _{3}}{3}\left[ 1-f(v,r)%
\right] ^{3}\right\} .  \label{Eqfr}
\end{equation}
Differentiating Eq. (\ref{Eqfr}) with respect to $v$ and using Eq. (\ref{sigma}), one finds:
\begin{equation}
\sigma =\frac{(n-2)}{2\kappa _{n}^{2}r^{n-2}}\dot{\varepsilon},
\label{sigma2}
\end{equation}
which shows that the functions $\rho(v,r)$ and $\sigma(v,r)$ are not arbitrary
for our ansatz metric (\ref{metric1}).

By solving Eq. (\ref{Eqfr}), the metric function is found to be:
\begin{equation}
f(v,r)=1+\frac{\alpha _{2}r^{2}}{\alpha _{3}}\left\{ 1-\left( \sqrt{\gamma
+k^{2}(v,r)}+k(v,r)\right) ^{1/3}+\gamma ^{1/3}\left( \sqrt{\gamma
+k^{2}(v,r)}+k(v,r)\right) ^{-1/3}\right\} ,  \label{fr}
\end{equation}
where
\begin{eqnarray}
k(v,r) &=&\frac{1}{2}+\frac{3}{2}\gamma ^{1/3}+\frac{3\alpha
_{3}^{2}\varepsilon (v,r)}{2\alpha _{2}^{3}r^{n-1}},  \notag \\
\text{\ }\gamma &=&\left( \frac{\alpha _{3}^{2}-\alpha _{2}^{2}}{\alpha
_{2}^{2}}\right) ^{3}.  \label{kr}
\end{eqnarray}
The solution introduced by Eqs. (\ref{metric1}) and (\ref{kr}) is a
general spherically symmetric solution of third order Lovelock gravity
with the ansatz metric (\ref{metric1}) in the presence of a type II perfect
fluid, where $\rho(v,r)$, $\sigma(v,r)$ and $p(v,r)$ are related to each other
according to Eqs. (\ref{p}) and (\ref{sigma2}). This solution contains all the previous solutions of third order
Lovelock gravity introduced in the literature \cite{Deh1,Far} and contains some
new exact solutions which will be discussed in the rest of the paper.

\section{Static Solutions for Linear Equation of State:}\label{Stat}

In this section, we find the static solutions of third order Lovelock
gravity in the presence of a type I ($\sigma =0$) perfect fluid. Knowing the
equation of state, and using Eq. (\ref{p}), one may obtain the density
function $\rho (r)$, and therefore the energy function $\varepsilon (r)$
explicitly. For the linear equation of state $p=w\rho $, Eq. (\ref{p})
reduces to

\begin{equation}
\frac{d}{dr}\left[ r^{n-2}\rho (r)\right] +(n-2)wr^{n-3}\rho (r)=0,
\end{equation}
with the solution

\begin{equation}
\rho (r)=\frac{\lambda ^{2}}{r^{(w+1)(n-2)}},  \label{rhosol}
\end{equation}
where the integration constant $\lambda ^{2}$ is positive in order to have
the weak and dominant energy conditions. Using Eq. (\ref{mass}), the energy
function $\varepsilon (r)$ may be obtained as
\begin{eqnarray}
\varepsilon (r) &=&m-\frac{\lambda ^{2}}{[w(n-2)-1]r^{w(n-2)-1}};\hspace{%
0.5cm}w(n-2)\neq 1,  \notag \\
&=&m+\lambda ^{2}\ln r;\hspace{3.5cm}w(n-2)=1.  \label{mass1}
\end{eqnarray}

For $w=1$, with $q^{2}=\lambda ^{2}/(n-3)$,\ the function $k(r)$ becomes
\begin{equation}
k(r)=-1+\frac{3\alpha _{3}}{2\alpha _{2}^{2}}+\frac{3\alpha _{3}^{2}}{%
2\alpha _{2}^{3}}\left( \frac{m}{r^{n-1}}+\frac{q^{2}}{r^{2(n-2)}}\right) .
\label{k1}
\end{equation}
The solution given by Eqs. (\ref{fr}) and (\ref{k1}) is the asymptotically
flat static charged black hole of third order Lovelock gravity introduced in
\cite{Deh1}.

For $w=-1$ with the choice of $\lambda ^{2}=(n-1)\Lambda $, the function $%
k(r)$ reduces to
\begin{equation}
k(r)=-1+\frac{3\alpha _{3}}{2\alpha _{2}^{2}}+\frac{3\alpha _{3}^{2}}{%
2\alpha _{2}^{3}}\left( \Lambda +\frac{m}{r^{n-1}}\right) .  \label{k2}
\end{equation}
Due to the weak and dominant energy condition, $\Lambda $ should be positive
and therefore the solution given by Eqs. (\ref{fr}) and (\ref{k2}) presents
an asymptotically dS uncharged solution of third order Lovelock gravity.

\subsection{Black hole for $w=0:$}

\begin{figure}
\centering {\includegraphics[width=7cm]{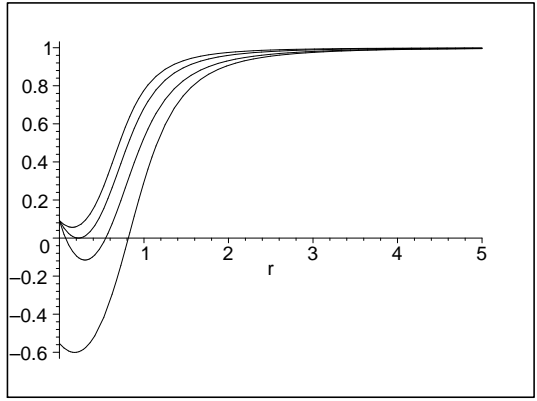} }
\caption{$f(r)$ vs. $r$ for $n=7$, $\protect\alpha _{2}=0.5$, $\protect%
\alpha _{3}=0.4$, $m =.1<\alpha_3/3$, $\lambda<\lambda_{\mathrm{ext}}$, $\lambda=\lambda_{\mathrm{%
ext}}$, $\lambda>\lambda_{\mathrm{ext}}$ and $m=.5>\alpha_3/3$ from up to down, respectively.}
\label{F1}
\end{figure}

The static case with $w=0$ gives a new asymptotically flat solution of third order Lovelock
gravity. The metric function $f(r)$ is the solution of the following
equation
\begin{equation}
r^{n-7}\left\{ r^{4}\left[ 1-f(r)\right] +\alpha _{2}r^{2}\left[ 1-f(r)%
\right] ^{2}+\frac{\alpha _{3}}{3}\left[ 1-f(r)\right] ^{3}\right\} =\lambda
^{2}r+m.  \label{Eqf0}
\end{equation}
The solution of Eq. (\ref{Eqf0}) is given in Eq. (\ref{fr})
with the following $k(r)$:
\begin{equation}
k(r)=-1+\frac{3\alpha _{3}}{2\alpha _{2}^{2}}+\frac{3\alpha _{3}^{2}}{%
2\alpha _{2}^{3}}\left( \frac{m}{r^{n-1}}+\frac{\lambda ^{2}}{r^{n-2}}%
\right) .  \label{k3}
\end{equation}
One may show that the above solution is asymptotically flat. In order to study the general structure of this solution, we first look for
the curvature singularities. It is easy to show that the Kretschmann scalar $%
R_{\mu \nu \lambda \kappa }R^{\mu \nu \lambda \kappa }$ diverges at $r=0$,
it is finite for $r\neq 0$ and goes to zero as $r\rightarrow \infty $. Thus,
there is an essential singularity located at $r=0$. Also, it is notable to
mention that the Ricci scaler is finite everywhere except at $r=0$, and
goes to zero as $r\rightarrow \infty $. The event horizon(s), if there
exists any, is (are) located at the root(s) of $g^{rr}=f(r)=0$:
\begin{equation}
r_{+}^{n-3}+\alpha _{2}r_{+}^{n-5}+\frac{\alpha _{3}}{3}r_{+}^{n-7}-\lambda
^{2}r_{+}-m=0.  \label{Hor}
\end{equation}
In order to find the number of positive real roots of Eq. (\ref{Hor}), one
should note that if the parameters $m$ and $\lambda $ are chosen such that
there exists a positive real root for $f(r_{\mathrm{ext}})=f^{\prime }(r_{%
\mathrm{ext}})=0$, then Eq. (\ref{Hor}) has one real root. Differentiating
Eq. (\ref{Eqf0}) with respect to $r$ and using $f(r_{\mathrm{ext}%
})=f^{\prime }(r_{\mathrm{ext}})=0$, one obtains:
\begin{equation}
(n-3)r_{\mathrm{ext}}^{n-4}+(n-5)\alpha _{2}r_{\mathrm{ext}}^{n-6}+\frac{%
\alpha _{3}}{3}(n-7)r_{\mathrm{ext}}^{n-8}-\lambda _{\mathrm{ext}}^{2}=0.
\label{Horext}
\end{equation}
One may find a relation between $m_{\mathrm{ext}}$ and $\lambda _{\mathrm{ext%
}}$\ for the case that $f(r)$ has only one real root by omitting $r_{\mathrm{%
ext}}$ between Eqs. (\ref{Horext}) and (\ref{Hor}) for $r_{+}=$ $r_{\mathrm{%
ext}}$. For $n=7$, the relation between $m_{\mathrm{ext}}$ and $\lambda _{%
\mathrm{ext}}$ is found to be
\begin{equation*}
\lambda _{\mathrm{ext}}^{2}=\frac{1}{9}\left( -6\alpha _{2}+6\sqrt{\alpha
_{2}^{2}+4\alpha _{3}-12m_{\mathrm{ext}}}\right) ^{1/2}\left( 2\alpha _{2}+%
\sqrt{\alpha _{2}^{2}+4\alpha _{3}-12m_{\mathrm{ext}}}\right) ,
\end{equation*}
which is real and positive provided $m<\alpha _{3}/3$. When $m<\alpha _{3}/3
$, the solution given by Eqs. (\ref{fr}) and (\ref{k3}) presents a naked
singularity if $\lambda <\lambda _{\mathrm{ext}}$, an extreme black hole for
$\lambda =\lambda _{\mathrm{ext}}$, and a black hole with inner and outer
horizons provided $\lambda >\lambda _{\mathrm{ext}}$. For $m\geq \alpha
_{3}/3$, the solution is a black hole with one event horizon. To be more
clear on this explanation, one may see the diagram of $f(r)$ versus $r$ for
these four cases in Fig. \ref{F1}.

The temperature may be obtained through the use of the definition of surface
gravity. One obtains:
\begin{equation}
T=\frac{f^{^{\prime }}(r_{+})}{4\pi }=\frac{(n-3)r_{+}^{n-4}+\alpha
_{2}(n-5)r_{+}^{n-6}+\frac{\alpha _{3}}{3}(n-7)r_{+}^{n-8}-\lambda ^{2}}{4\pi(r_{+}^{n-3}+2\alpha _{2}r_{+}^{n-5}+r_{+}^{n-7}\alpha _{3})}.
\label{temperature}
\end{equation}
The entropy of asymptotically flat black holes of Lovelock gravity is \cite
{Myer}
\begin{equation}
S=\frac{2\pi}{\kappa_n^2}\sum_{p=1}^{[(n-1)/2]}p\alpha _{k}^{\prime }\int d^{n-2}x\sqrt{%
\tilde{g}}\tilde{\mathcal{L}}_{p-1}  \label{Ent}
\end{equation}
where the integration is done on the $(n-2)$-dimensional spacelike
hypersurface of Killing horizon, $\tilde{g}_{\mu \nu }$ is the induced
metric on it, $\tilde{g}$ is the determinant of $\tilde{g}_{\mu \nu }$ and $%
\tilde{\mathcal{L}}_{k}$ is the $k$th order Lovelock Lagrangian of $\tilde{g}%
_{\mu \nu }$. It is a matter of calculation to obtain the entropy as
\begin{equation}
S=\frac{2\pi V_{n-2}}{\kappa_n^2}r_{+}^{n-2}\left( 1+\frac{2(n-2)\alpha _{2}}{%
(n-4)r_{+}^{2}}+\frac{(n-2)\alpha _{3}}{(n-6)r_{+}^{4}}\right),
\label{Ent2}
\end{equation}
where $V_{n-2}$ is the volume of $(n-2)$-dimensional unit sphere.
Using Eqs. (\ref{temperature}) and (\ref{Ent2}), one may calculate the integral of $\int{TdS}$
\begin{equation}
\int{T(r_+)\frac{\partial S}{\partial r_+}dr_+}=\frac{(n-2)V_{n-2}}{2\kappa_n^2}m,
\end{equation}
which is proportional to the geometrical mass $m$.

Here, it is worth to consider a peculiar property of the above solutions
which does not occur in the lower order Lovelock gravity.
The solution with vanishing geometrical mass $m=0$ presents a black hole with
two inner and outer horizons provided $\lambda > \lambda_{\mathrm{ext}}$, where
$\lambda_{\mathrm{ext}}$ can be obtained through the use of Eqs. (\ref{Hor}) and (\ref{Horext}) as:
\begin{eqnarray}
&& \lambda _{\mathrm{ext}}^{2}=r_{\mathrm{ext}}^{n-4}\left(r_{\mathrm{ext}}^{4}+\alpha _{2}r_{\mathrm{ext}}^{2}+\frac{%
\alpha _{3}}{3}\right),\\
&& r_{\mathrm{ext}}^{2}=-\frac{(n-6)}{2(n-4)}\left\{-\alpha_2+\sqrt{\alpha_2^2-\frac{4(n-4)(n-8)}{3(n-6)^2}\alpha_3}\right\}. \label{rexm0}
\end{eqnarray}
The radius of the extreme black hole is zero for lower order Lovelock gravity and $n=8$ as one may see from Eq. (\ref{rexm0}),
and therefore the solution in these cases presents a black hole with only one horizon if $\lambda >0$. However,
the solution for $n=7$ with positive $\alpha_3$ or $n>8$ with negative $\alpha_3$
presents a black hole with two inner and outer horizons if $\lambda>\lambda _{\mathrm{ext}}$.
This is a peculiar feature of third order Lovelock gravity which does not occur in
the lower order Lovelock gravity in the presence of a perfect fluid.
\subsection{Black hole for $w=(n-2)^{-1}:$}
\begin{figure}
\centering {\includegraphics[width=7cm]{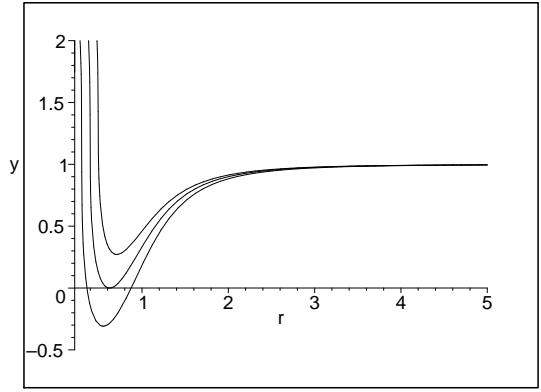} }
\caption{$f(r)$ vs. $r$ for $n=8$, $\protect\alpha _{2}=0.5$, $\protect%
\alpha _{3}=0.3$, $\protect\lambda =.5$, $m<m_{\mathrm{ext}}$, $m=m_{\mathrm{%
ext}}$, and $m>m_{\mathrm{ext}}$ from up to down, respectively.}
\label{F2}
\end{figure}

As in the case of $w=0$, the static case with $w=(n-2)^{-1}$ gives a new
solution of third order Lovelock gravity with
\begin{equation}
k(v,r)=\frac{1}{2}+\frac{3}{2}\gamma ^{1/3}+\frac{3\alpha _{3}^{2}(m+\lambda
^{2}\ln r)}{2\alpha _{2}^{3}r^{n-1}},  \label{kln}
\end{equation}
which is asymptotically flat. Again, the Kretschmann scalar $R_{\mu \nu \lambda \kappa }R^{\mu \nu \lambda
\kappa }$ diverges at $r=0$, it is finite for $r\neq 0$ and goes to zero as $%
r\rightarrow \infty $. Thus, there is an essential singularity located at $%
r=0$. The event horizon(s), if there exists any, is (are) located at the
root(s) of:
\begin{equation}
r_{+}^{n-3}+\alpha _{2}r_{+}^{n-5}+\frac{\alpha _{3}}{3}r_{+}^{n-7}-m-%
\lambda ^{2}\ln r_{+}=0.
\end{equation}
Following our leading procedure in the above subsection for $n=7$, the
relation between $m_{\mathrm{ext}}$ and $\lambda _{\mathrm{ext}}$ is found
to be
\begin{equation*}
m_{\mathrm{ext}}=\frac{\alpha _{3}}{3}+\frac{\lambda _{\mathrm{ext}}^{2}}{4}+%
\frac{\alpha _{2}}{8}\left( \sqrt{\alpha _{2}^{2}+4\lambda _{\mathrm{ext}%
}^{2}}-\alpha _{2}\right) ^{1/2}-\frac{\lambda _{\mathrm{ext}}^{2}}{2}\ln
\left( \frac{-\alpha _{2}+\sqrt{\alpha _{2}^{2}+4\lambda _{\mathrm{ext}}^{2}}%
}{4}\right) .
\end{equation*}
The solution given by Eqs. (\ref{fr}) and (\ref{kln}) presents a naked
singularity if $m<m_{\mathrm{ext}}$, an extreme black hole for $m=m_{\mathrm{%
ext}}$, and a black hole with inner and outer horizons provided $m>m_{%
\mathrm{ext}}$. The diagrams of $f(r)$ versus $r$ for these cases are shown in
Fig. \ref{F2}. Again, one may note that $\int{TdS}$ is proportional to the geometrical mass of the black hole.

\subsection{Strange quark matter:}

For completeness, we consider the linear equation of state of a strange quark
matter (SQM). This is due to the fact that near the singularity, matter may
be in the highest known density form, which brings in the SQM. In 4 dimensions,
the SQM fluid is characterized by the equation of state $p=w(\rho -4B/3)$ where $B$ is the bag constant indicating
the difference between the energy density of the perturbative and nonperturbative QCD vacuum \cite{SQM}.
Here, we consider the equation of state $p=w(\rho -4B)$ as
a generalization of equation of state of SQM in $n$ dimensions, where $B$ is a constant \cite{Har,Gosh}. In this case, the functions $%
\rho (r)$ and $\varepsilon(r)$ reduce to:
\begin{equation}
\rho (r)=\frac{\lambda ^{2}}{r^{(w+1)(n-2)}}+\frac{B}{w+1},
\end{equation}
\begin{eqnarray}
\varepsilon (r) &=&m+\frac{4wBr^{n-1}}{(1+w)(n-1)}-\frac{\lambda ^{2}}{%
[(n-2)w-1]r^{(n-2)w-1}};\hspace{0.3cm}w(n-2)\neq 1,  \notag \\
&=&m+\frac{4wBr^{n-1}}{(1+w)(n-1)}+\lambda ^{2}\ln r;\hspace{3cm}(n-2)w=1,
\label{mass3}
\end{eqnarray}
respectively. The function $k(r)$ for the above two cases becomes
\begin{eqnarray}
k(r) &=&-1+\frac{3\alpha _{3}}{2\alpha _{2}^{2}}+\frac{6wB\alpha _{3}^{2}}{%
\alpha _{2}^{3}(1+w)(n-1)}+\frac{3\alpha _{3}^{2}}{2\alpha _{2}^{3}}\left(
\frac{m}{r^{n-1}}+\frac{\lambda ^{2}}{r^{(1+w)(n-2)}}\right); \hspace{0.3cm}w(n-2)\neq 1 ,  \notag \\
&=&-1+\frac{3\alpha _{3}}{2\alpha _{2}^{2}}+\frac{6wB\alpha _{3}^{2}}{\alpha
_{2}^{3}(1+w)(n-1)}+\frac{3\alpha _{3}^{2}\left( m+\lambda ^{2}\ln r\right)
}{2\alpha _{2}^{3}r^{n-1}}; \hspace{1cm}(n-2)w=1.  \label{k4}
\end{eqnarray}
The $B$-term in the above expressions guarantees that the solution
is asymptotically (A)dS even in the cases of $w=1$, $w=0$ and $w=(n-2)^{-1}$, which are
asymptotically flat for normal matter.

\section{Radiating solutions:}\label{Rad}

Now, we want to find the radiating solutions of third order Lovelock gravity
in the presence of energy-momentum tensor (\ref{EMten}) for linear equation of
state with $\sigma (v,r)$ given in Eq. (\ref{sigma}). In this case, the
energy density $\varepsilon (v,r)$ becomes
\begin{eqnarray}
\varepsilon (v,r) &=&m(v)-\frac{\lambda^{2}(v)}{[w(n-2)-1]r^{w(n-2)-1}};\hspace{%
0.5cm}w(n-2)\neq 1.  \notag \\
&=&m(v)+\lambda ^{2}(v)\ln r;\hspace{2cm}w(n-2)=1.  \label{mass4}
\end{eqnarray}
When $w=-1$, the solution is the asymptotically (A)dS uncharged Vaidya-type
solution introduced in Ref. \cite{Far}.

For $w=1$ with $q^2(v)=\lambda^2(v)/(n-3)$, the solution is a new
solution which may be called the charged Vaidya-type solution of third order
Lovelock gravity for which the metric function is given by Eq. (\ref{fr})
with:
\begin{equation*}
k(v,r)=\frac{1}{2}+\frac{3}{2}\gamma ^{1/3}+\frac{3\alpha _{3}^{2}}{2\alpha
_{2}^{3}}\left\{ \frac{m(v)}{r^{n-1}}+\frac{q^{2}(v)}{r^{2(n-2)}}\right\} .
\end{equation*}
In this case, the energy-momentum tensor is the sum of $\sigma v_{\mu} v_{\nu}$ and the Maxwell
energy-momentum tensor
\begin{equation}
T_{\mu \nu }^{(em)}=F_{\mu \lambda }F_{\nu }^{\text{ \ }\lambda }-\frac{1}{4}%
g_{\mu \nu }F_{\rho \sigma }F^{\rho \sigma },
\end{equation}
for a point charge $q(v)$ with electromagnetic field tensor $F_{\mu \nu
}=\partial _{\mu }A_{\nu }-\partial _{\nu }A_{\mu }$ and the following
potential:
\begin{equation*}
A_{\mu }=\frac{q(v)}{r^{n-2}}\delta _{\mu }^{t}.
\end{equation*}
Using Eq. (\ref{sigma2}), one obtains
\begin{equation}
\sigma =\frac{(n-2)}{2\kappa _{n}^{2}}\left( \frac{\dot{m}}{r^{n-2}}-%
\frac{2q\dot{q}}{r^{2n-5}}\right) ,  \label{sigma3}
\end{equation}
which shows that when $\dot{q}<0$, the weak and dominated energy conditions
will be satisfied. While for $\dot{q}\geq 0$, the energy conditions is
satisfied for $r>r_{c}$, where $r_{c}=(2q\dot{q}/\dot{m})^{1/(n-3)}$. But in
realistic situations the particle cannot get into the region $r<r_{c}$
because of the Lorentz force, and therefore the weak and dominant energy
conditions are satisfied for the charged Vaidya-type solution \cite{Ori}. In
the limit of $\alpha _{2}=\alpha _{3}=0$, this solution reduces to the
charged Vaidya solution of Einstein gravity introduced in Ref. \cite{Sull}.

The nature of the singularity (to be naked or hidden) can be characterized
by the existence of radial null geodesics coming out of the singularity. The
nature of the singularity is exactly the same as the uncharged solutions
explained in Ref. \cite{Far}. Thus, we only compare the strength of the
singularity with the case of uncharged solution. The solution satisfies a
strong curvature condition (SCC) \cite{scc} or limiting focusing condition (LFC)
\cite{lfc} provided the limits of $\tau ^{2}\Phi $ or $\tau \Phi $ are positive,
respectively, where $\tau $ is an affine parameter and $\Phi $ is
\begin{equation*}
\Phi \equiv R_{\mu \nu }v^{\mu }v^{\nu }.
\end{equation*}
Using the fact that $dr/d\tau =(dv/d\tau )f/2$, one can show that
\begin{equation}
\Phi =-\frac{2(n-2)\dot{f}}{rf^{2}}\left( \frac{dr}{d\lambda }\right) ^{2},
\label{Phi}
\end{equation}
and the radial null geodesic satisfies the differential equation
\begin{equation}
\frac{d^{2}r}{d\lambda ^{2}}\simeq \frac{2\dot{f}}{f^{2}}\left( \frac{dr}{%
d\lambda }\right) ^{2}.  \label{d2r}
\end{equation}
Now, we consider the strength of the singularity for the following two cases:

1. $m(v)=m_{0}\theta(v)v^{n-3}$ and $q^{2}(v)=q_{0}^{2}\theta(v)v^{2(n-3)}$, where $m_{0}$ and
$q_{0}$ are two arbitrary constants, and $\theta(v)$ is the step function which is $1$
for $v>0$ and zero for $v<0$: In this case, the limits of $f$ and
$\dot{f}$\ are $1$ and zero as $r=v\rightarrow 0$, respectively. Thus, one
finds that the limit of $\tau \Phi $ and $\tau ^{2}\Phi $ are zero as $\tau $
goes to zero, and therefore neither the SCC nor LFC are satisfied along a
radial null geodesic. The uncharged solution satisfies the LFC \cite{Far},
while the charged solution does not satisfy LFC, and therefore charge
weakens the strength of the singularity.

2. $m(v)=m_{0}\theta(v)v^{n-4}$ and $q^{2}(v)=q_{0}^{2}\theta(v)v^{2n-7}$, where again $m_{0}$
and $q_{0}$ are two arbitrary constants: In this case, the limits of $f$
and $\dot{f}$  as $r$ and $v$ go to zero are:
\begin{eqnarray}
\underset{r\rightarrow 0}{\lim }f &=&1  \label{limf} \nonumber \\
\underset{r\rightarrow 0}{\lim }\overset{.}{f} &=&-\frac{(3\alpha
_{3}^{2})^{1/3}}{3}\frac{\left[ (n-4)m_{0}-(2n-7)q_{0}^{2}\right] )}{%
(m_{0}-q_{0}^{2})^{2/3}}\equiv \dot{f}_{0}  \label{limfdot}
\end{eqnarray}
Using Eqs. (\ref{d2r}) and (\ref{limfdot}), one may show that the radial null
geodesic near $r=v=0$ is $r\simeq (\dot{f}_{0})^{-1}\ln (\tau +1)$. Using
this result and Eqs. (\ref{Phi}) and (\ref{limfdot}),
one finds that the limit of $\tau \Phi $ is positive while the limit of $%
\tau ^{2}\Phi $ is zero as $\tau $ goes to zero, and therefore as in the
case of uncharged solution \cite{Far}, only LFC is satisfied along a radial
null geodesic. Although the radiating solution with $w=0$ is a new solution
with
\begin{equation*}
k(r)=-1+\frac{3\alpha _{3}}{2\alpha _{2}^{2}}+\frac{3\alpha _{3}^{2}}{%
2\alpha _{2}^{3}}\left( \frac{m(v)}{r^{n-1}}+\frac{\lambda^2 (v)}{r^{n-2}}%
\right) ,
\end{equation*}
but similar calculations show that the strength of the singularity for
this case is exactly the same as when $\lambda (v)=0$ \cite{Far}, and therefore we will
not present them here.

As in the case of static solutions, the effect of using the equation of
state of SQM instead of normal matter is to make the solutions
asymptotically (A)dS. These solutions reduce to the radiating solutions
of Einstein gravity in the presence of SQM introduced in Ref. \cite{Gosh}.

\section{ CLOSING REMARKS}
We considered the third order Lovelock gravity for a spherically symmetric spacetime
in the presence of a type II perfect fluid. Due to the fact that
the ansatz metric (\ref{metric1}) had only one unknown function, the
nonvanishing components of $\mathcal{G}_{\mu \nu }$ of
the LHS of Eq. (\ref{field}) were
related to each other according to the following equations:
\begin{eqnarray*}
\mathcal{\dot{G}}_{v }^{\text{ \ }v } &=&\frac{1}{r^{n-2}}\frac{\partial
}{\partial r}(r^{n-2}\mathcal{G}_{v}^{\text{ \ }r}), \\
\mathcal{G}_{i}^{\text{ \ }i} &=&\frac{1}{(n-2)r^{n-3}}\frac{\partial }{%
\partial r}(r^{n-2}\mathcal{G}_{v }^{\text{ \ }v}).
\end{eqnarray*}
Thus in our analysis, the functions $\rho$, $p$ and $\sigma$
in the energy-momentum tensor (\ref{EMten}) were not arbitrary.
Nevertheless, we found a general solution for linear equation of state which contains all the
known solutions of third order Lovelock gravity, and also some new
static and radiating solutions.
We investigated the properties of static solutions for $w=0$
and $w=(n-2)^{-1}$ with zero and nonzero $B$, and found that these
solutions may be interpreted as black holes with
inner and outer horizons, extreme black holes, naked singularities
or black holes with one horizon provided the metric parameters are chosen suitable.
We found that the solution with $w=0$ and $m=0$ may present a black hole
with inner and outer horizons. This is a peculiar feature of
third order Lovelock gravity which does not occur in
lower order Lovelock gravity in the presence of perfect fluid.
We also presented the new radiating solutions for $w=0$ and $w=(n-2)^{-1}$,
and compared the singularity strengths of these solutions and the known
radiating solutions of third order Lovelock gravity in the literature.
Although naked singularity is inevitably formed in
third order Lovelock gravity for the general energy-momentum tensor (\ref{EMten}), the strength of the singularity
depends on the rate of increase of $\varepsilon(v,r)$ with respect to $v$. For the case of charged solutions
when the rate of increase of charge is large enough, then the strength of the singularity
is weaker than that of the uncharged case. That is, the charge with a suitable rate of increase
weakens the strength of the singularity.
We also found that the presence of SQM instead of normal matter
changes the asymptotic behavior of the solutions from flat to (A)dS.

Although we investigated the solutions for $w=0$, $w=1$ and $w=(n-2)^{-1}$, one may
consider the solutions for other values of $w$.
If one likes to obtain the most general spherically symmetric solution of
Lovelock gravity in the presence of a
type II perfect fluid with arbitrary $\rho$, $p$ and $\sigma$,
then one should consider the following metric:
\[
ds^{2}=-A(v,r)dv^{2}+2B(v,r)drdv+r^{2}C(v,r) \gamma_{i j}d\theta^i\theta^j.
\]
\textbf{Acknowledgements}

This work has been supported by Research Institute for Astrophysics and
Astronomy of Maragha.

\end{document}